\documentclass[12pt]{iopart}
\usepackage{graphicx} 
\usepackage{flexisym}
\usepackage{textcomp,gensymb}
\usepackage{booktabs,multirow}
\usepackage{makecell}
\usepackage{natbib}

\usepackage{setspace}

\newcolumntype{P}[1]{>{\centering\arraybackslash}p{#1}}
\newcolumntype{L}[1]{>{\raggedright\arraybackslash}p{#1}}
\begin{document}

\title[Development of LAF-TMS]{Development of Linear Astigmatism Free - Three Mirror System (LAF-TMS)}

\author{Woojin Park$^1$, Seunghyuk Chang$^2$, Jae Hyuk Lim$^3$, Sunwoo Lee$^1$, Hojae Ahn$^4$, Yunjong Kim$^5$, Sanghyuk Kim$^5$, Arvid Hammar$^6$, Byeongjoon Jeong$^7$, Geon Hee Kim$^7$, Hyoungkwon Lee$^8$, Dae Wook Kim$^{9,10}$, and Soojong Pak$^{1,4,*}$}

\address{$^1$ School of Space Research and Institute of Natural Science, Kyung Hee University, Yongin 17104, Republic of Korea}
\address{$^2$ Center for Integrated Smart Sensors, Daejeon 34141, Republic of Korea}
\address{$^3$ Department of Mechanical Engineering, Chonbuk National University, Jeonju 54896, Republic of Korea}
\address{$^4$ Department of Astronomy and Space Science, Kyung Hee University, Yongin 17104, Republic of Korea}
\address{$^5$ Korea Astronomy and Space Science Institute, Daejeon 34055, Republic of Korea}
\address{$^6$ Omnisys Instruments AB,V\"{a}stra Fr\"{o}lunda, SE-421 32, Sweden}
\address{$^7$ Korea Basic Science Institute, 169-148, Daejeon 34133, Republic of Korea}
\address{$^8$ Green Optics, Cheongju 28126, Republic of Korea}
\address{$^9$ James C. Wyant College of Optical Sciences, University of Arizona, Tucson, Arizona 85721, USA}
\address{$^{10}$ Department of Astronomy and Steward Observatory, University of Arizona, Tucson, Arizona 85719, USA}

\eads{\mailto{woojinpark@khu.ac.kr}, \mailto{soojong@khu.ac.kr, $^*$Corresponding author}}
\vspace{10pt}
\begin{indented}
\item[]November 2019
\end{indented}

\doublespacing
\begin{abstract}
We present the development of Linear Astigmatism Free - Three Mirror System (LAF-TMS). This is a prototype of an off-axis telescope that enables very wide field of view (FoV) infrared satellites that can observe Paschen-$\alpha$ emission, zodiacal light, integrated star light, and other infrared sources. It has the entrance pupil diameter of 150 mm, the focal length of 500 mm, and the FoV of 5.5$\degree$ $\times$ 4.1$\degree$. LAF-TMS is an obscuration-free off-axis system with minimal out-of-field baffling and no optical support structure diffraction. This optical design is analytically optimized to remove linear astigmatism and to reduce high-order aberrations. Sensitivity analysis and Monte-Carlo simulation reveal that tilt errors are the most sensitive alignment parameters that allow $\sim$1$^\prime$. Optomechanical structure accurately mounts aluminum mirrors, and withstands satellite-level vibration environments. LAF-TMS shows optical performance with 37 $\mu$m FWHM of the point source image satisfying Nyquist sampling requirements for typical 18 $\mu$m pixel Infrared array detectors. The surface figure errors of mirrors and scattered light from the tertiary mirror with 4.9 nm surface micro roughness may affect the measured point spread function (PSF). Optical tests successfully demonstrate constant optical performance over wide FoV, indicating that LAF-TMS suppresses linear astigmatism and high-order aberrations. 
\end{abstract}
\noindent{\it Keywords\/}: Astronomical instrumentation (799), Optical telescopes (1174), Reflecting telescopes (1380), Wide-field telescopes (1800)

\maketitle
\section{Introduction}
All sky survey missions in infrared wavelength are important in understanding the early universe. Infrared observations are generally performed in space because the earth's atmosphere absorbs infrared light. Infrared all-sky surveys first began with the Infrared Astronomical Satellite (IRAS) \citep{neugebauer1984}. It discovered distant galaxies, intergalactic cirrus, planetary disks, and many asteroids \citep{houck1984,minorplanet}. Since then, several infrared satellites have been developed. Infrared Array Camera (IRAC) in the Spitzer space telescope has observed high-z galaxies with a four channel camera that covers 3.6, 4.5, 5.8, and 8.0 $\mu$m \citep{fazio2004}. Interstellar medium, star formation, planetary disks studies, and formation and evolution of galaxies are prime scientific subjects in near- and mid- infrared wavelength \citep{onaka2007}. Infrared Camera (IRC) for the Akari satellite observed these targets in the spectral range of 1.8 - 26.5 $\mu$m \citep{ishihara2010}. The Wide-Field Infrared Survey Explorer (WISE) is another all-sky infrared satellite that observes in four infrared channels (3.3, 4.7, 12, 23 $\mu$m) \citep{duval2004}.

The main optics for most infrared cameras, including four satellites introduced above, adapt on-axis reflective mirrors \citep{werner2012,mainzer2005}. This optical system, however, is limited to narrow field of view (FoV) observations since the wider the FoV observations, the larger the secondary mirrors become, resulting in serious obscuration. The alternative is to use refractive optical system. Multi-purpose Infra-Red Imaging System (MIRIS) uses five refractive lenses, and the system covers 3.67$\degree$ $\times$ 3.67$\degree$ FoV in the wavelength coverage from 0.9 to 2.0 $\mu$m. It observes Paschen-$\alpha$ emission lines along the Galactic plane and the cosmic infrared background (CIB) \citep{ree2010,han2014}. However, observable wavelength bands are highly limited due to availability of lens materials.

Classical off-axis design alleviates wavelength limitations and avoids the obscuration problem, but it still faces limitations for wide FoV observations due to linear astigmatism. Linear astigmatism is a dominant aberration of classical off-axis telescopes, and it significantly degrades image quality, especially for large FoV systems \citep{chang2005,chang2016}. Near-infrared Imaging Spectrometer for Star formation history (NISS) reduces linear astigmatism by putting additional relay-lenses for wide FoV observations \citep{moon2018}.

However, linear-astigmatism-free confocal off-axis reflective system overcomes both the FoV and wavelength limits without the need for correcting lenses. Confocal off-axis design whose optical components share focuses instead of sharing axis can eliminate linear astigmatism by properly selecting mirror surface parameters and tilt angles \citep{chang2006}. Schwarzschild-Chang off-axis telescope is the first telescope with a linear-astigmatism-free two mirror system. \cite{kim2010} verified the feasibility of linear-astigmatism-free three-mirror optical design. \citet{chang2015} extended his theory to N-conic mirror system, which enables building a Linear Astigmatism Free - Three Mirror System (LAF-TMS).

We introduce a prototype LAF-TMS telescope for wide FoV infrared satellites for all-sky surveys. Optical design of the linear-astigmatism-free system is described in Section~\ref{sec:opticdesign}. Section~\ref{sec:tor} explores system tolerance and sensitivity of each component. Freeform aluminum mirror specification and manufacturing process are discussed in Section~\ref{sec:mirror}. Section~\ref{sec:optomechanics} presents optomechanical design and finite element analysis results. Finally, optical performance of the actual LAF-TMS is examined in Section~\ref{sec:perform}. 

\section{Optical design}
\label{sec:opticdesign}
\begin{figure}[!ht]
\centering\includegraphics[width=8.3cm]{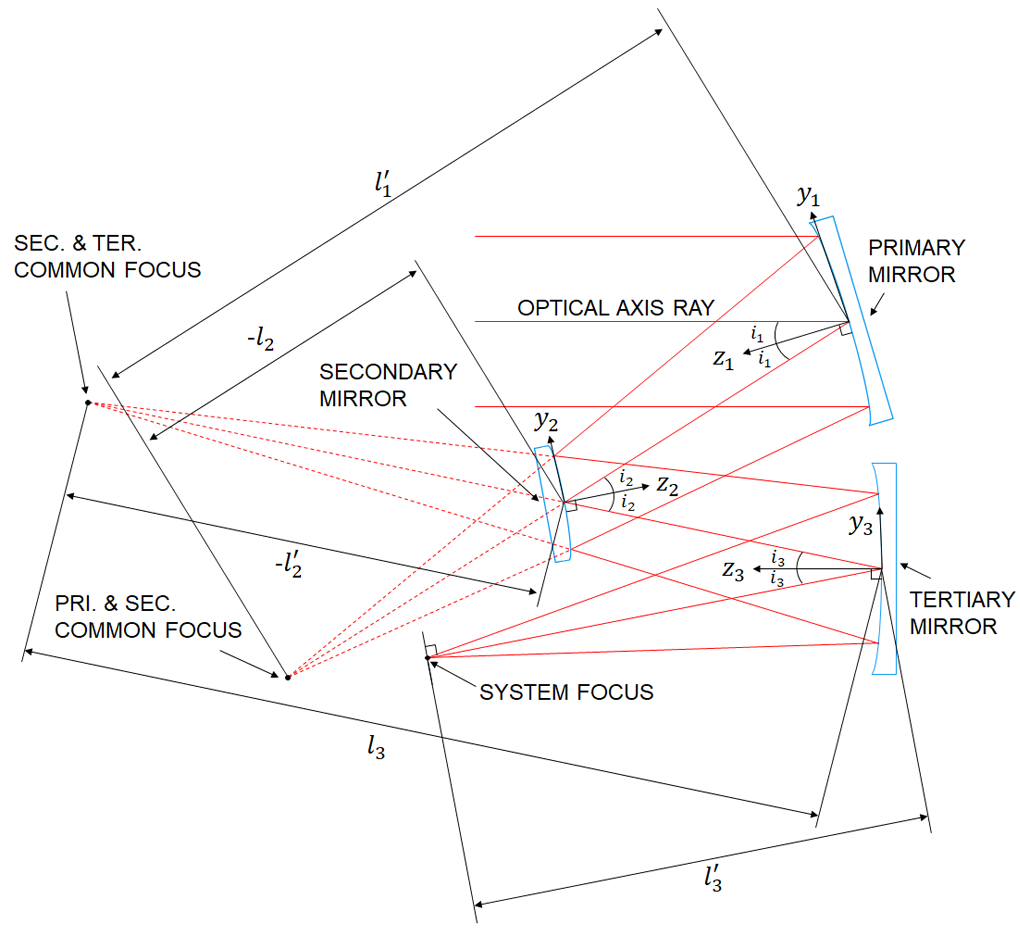}
\caption{\label{opticaldesign}The optical layout of LAF-TMS. Optical path is drawn in red solid lines.}
\end{figure}

LAF-TMS is a linear-astigmatism-free confocal off-axis three mirror telescope. Figure~\ref{opticaldesign} illustrates the optical layout of LAF-TMS, where optical path is indicated by red-solid lines. The mirror surface combination of the base confocal off-axis system is a parabolic concave primary mirror (M1), an ellipsoidal convex secondary mirror (M2), and an ellipsoidal concave tertiary mirror (M3). Thus, M1 shares its focus with M2. Two M2 focuses are shared with M1 and M3, respectively. One of the M3 focuses is shared with M2, and the other one is the system focus. The M1 \& M2 common focus and the M2 \& M3 common focus are labeled in Figure~\ref{opticaldesign}.

Mirror tilt angles and inter mirror distances, also called despace, are accurately calculated to satisfy the linear-astigmatism-free condition, as in expressed in equation (\ref{eq1}) \citep{chang2013}:
\small
\begin{eqnarray}
\frac{l^'_2}{l_2}\frac{l^'_3}{l_3}\tan i_1 + \left(1 + \frac{l^'_2}{l_2} \right)\frac{l^'_3}{l_3} \tan i_2 + \left(1 + \frac{l^'_3}{l_3} \right)\tan i_3 = 0 \label{eq1}
\end{eqnarray}
\normalsize
In equation (\ref{eq1}), $i_{1,2,3}$ are tilt angles of each mirror, and $l_{2,3}$ and $l'_{2,3}$ are the front and the back focal lengths of each mirror, respectively, as denoted in Figure~\ref{opticaldesign}.	
The calculated optical parameters for the prototype LAF-TMS are listed in Table~\ref{tab:optic}.

\begin{table}[!ht]
\centering
\caption{Optical parameters of LAF-TMS}
\label{tab:optic}
\footnotesize
\begin{tabular}{L{2.5cm} L{2cm}}
\br
Parameter           & Value\\
\mr
$l_{2}$	            & 625 mm \\
$l_{3}$             & 781.19 mm \\
$l'_{3}$            & 413.50 mm \\
$i_{1}$	            & 16\degree \\
$i_{2}$	            & 22\degree \\
$i_{3}$	            & 11.38\degree \\
EPD                 & 150 mm \\
Focal length 	    & 500 mm \\
Field of View 	    & 5.51$\degree$ $\times$ 4.13$\degree$ \\
\br
\end{tabular}
\end{table}

The entrance pupil diameter (EPD) is 150 mm, and the focal length is 500 mm. LAF-TMS has a wide FoV of 5.51$\degree$ $\times$ 4.13$\degree$ when used with a charge-coupled device (CCD) camera with a 6 $\mu$m 8716 $\times$ 6132 (or size of 49 $\times$ 36.7 mm) format sensor (ML50100, \cite{fli}). Aperture stop is located at the M2 surface to compensate for the mirror size of M1 and M3. Each mirror surface is optimized to reduce the higher-order aberrations while simultaneously satisfying the linear-astigmatism-free property. The resulting mirror shapes are freeform \citep{chang2019}.

\begin{figure}[!ht]
\centering\includegraphics[width=6cm]{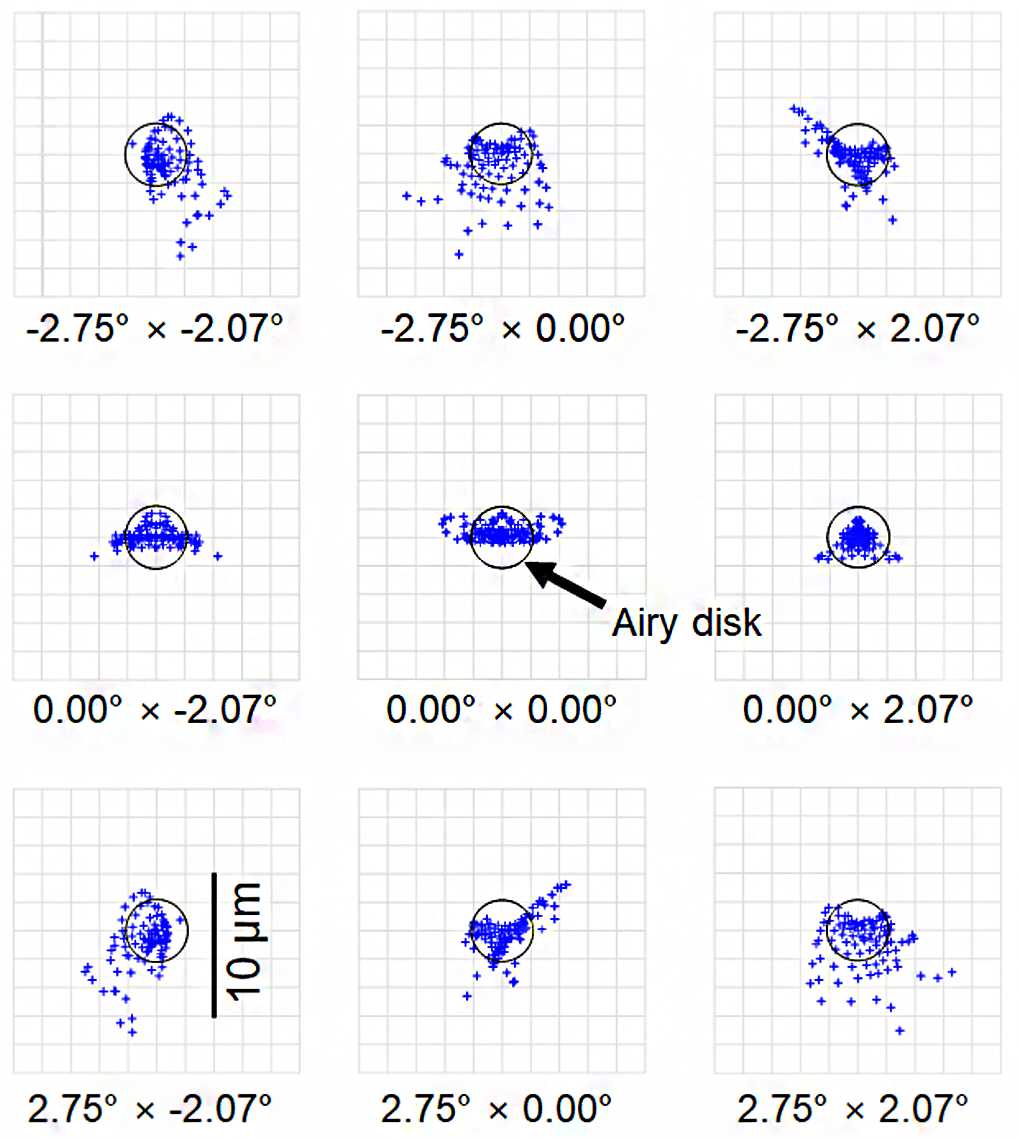}
\caption{\label{spotdiagram}A spot diagram of the confocal off-axis LAF-TMS design. Airy disks for 0.532 $\mu$m wavelength are shown as black circles.}
\end{figure}

The system targets for the infrared camera. However, the optical design satisfies diffraction limited performance in 0.532 $\mu$m wavelength because we perform conservative performance tests in visible wavelength (Figure~\ref{spotdiagram}). An airy disk diameter in 0.532 $\mu$m wavelength is 2.16 $\mu$m. The spot diagrams show that an excellent performance is obtained over a full FoV due to zero linear astigmatism and small higher order aberrations.
\section{Tolerance analysis}
\label{sec:tor}
Optical performance degradation due to manufacturing errors is evaluated by tolerance and sensitivity analysis \citep{wang2013}. \citet{kim2010} defined tolerance parameters and coordinate system for tolerance analysis. Despace indicates inter-mirror distance, while in-plane movements of mirror surfaces are expressed in x- and y-decenters. The mirror offset towards surface normal is defined as z-decenter. CODE V and ZEMAX are used for sensitivity analysis and Monte-Carlo simulation, respectively.

\subsection{Sensitivity analysis}
\label{sec:sensitivity}
\begin{figure*}[t]
\centering\includegraphics[width=14cm]{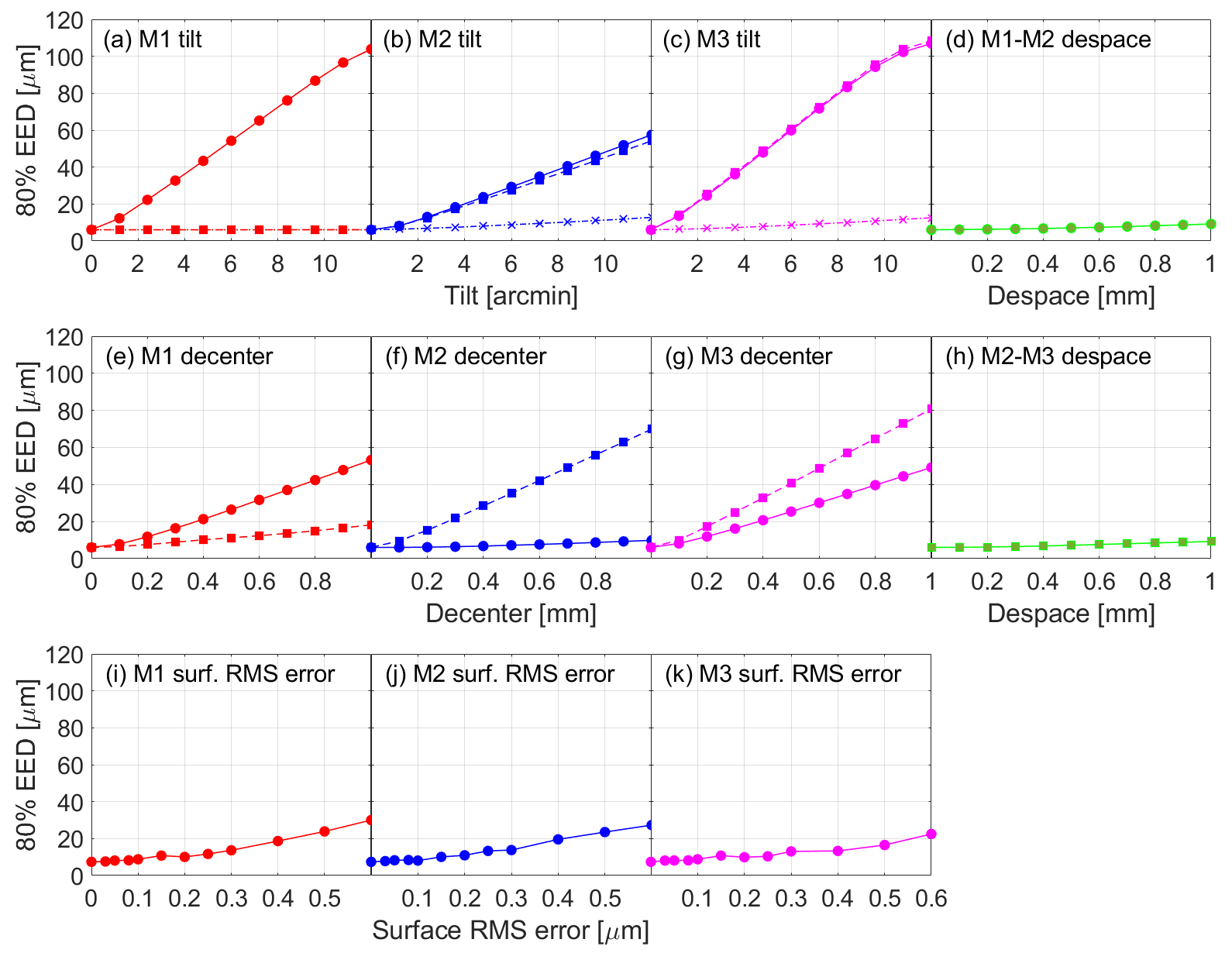}
\caption{\label{surfacerms}Sensitivity analysis results of M1 (red), M2 (blue), and M3 (magenta): (a - c) $\alpha$- (circle), $\beta$- (square), and $\gamma$- (cross) tilts, (e - g) x- (circle), and y- (square) decenters, (i - k) surface RMS errors, and (d) M1-M2 (circle), (h) M2-M3 (square) despaces. $\gamma$-tilt of M1 overlaps with its $\beta$-tilt, and $\alpha$-tilt of M3 also overlaps with its $\beta$-tilt.}
\end{figure*}

We performed sensitivity analysis on each surface's tilt, decenter, despace, and root mean square (RMS) error. The criterion of the sensitivity analysis is the 80 $\%$ encircled energy diameter (EED) for the point source with 0.532 $\mu$m wavelength. Sensitivities are calculated at five field angles, i.e., [$\alpha$ = -2.75$\degree$, $\beta$ = -2.07$\degree$], [-1.38$\degree$, -1.03$\degree$], [0.00$\degree$, 0.00$\degree$], [1.38$\degree$, 1.03$\degree$], and [2.75$\degree$, 2.07$\degree$]. The mean values of 80 $\%$ EED from the five fields are taken for overall performance variation to decide tolerance limits of Monte-Carlo simulation \citep{lee2010}. Figure~\ref{surfacerms} summarizes the sensitivity analysis results. 

M1, M2, and M3 are indicated in red, blue, and magenta, respectively. Calculated EED results of negative and positive tolerances are symmetry. Analysis results show that despaces, $\gamma$- tilts for all three mirrors, $\beta$- tilt, y-decenter of M1, and the x-decenter of M2 are practically insensitive parameters, which highlight the robustness of the LAF-TMS design solution. By considering mechanical fabrication tolerances,  $\alpha$- and $\beta$- tilts are the most sensitive parameters. Decenter is less sensitive compared to tilt as we often assemble and align optical components within $\pm$0.1 mm tolerances. M3 is slightly less sensitive than the other mirrors in terms of the surface RMS error.

\subsection{Monte-Carlo simulation}
Monte-Carlo method is the most common method for a statistical system tolerance analysis that simulates the comprehensive performance with the errors altogether \citep{burge2010,funck2010,kus2017}. Tolerance parameters are despace, decenter, and tilt. A focal position is set to the compensator. Detailed tolerances used for the Monte-Carlo simulation are listed in Table~\ref{tab:tor}. 

\begin{table}[!ht]
\centering
\caption{Tolerance parameters for the LAF-TMS Monte-Carlo simulation}
\label{tab:tor}
\footnotesize
\begin{tabular}{L{3.5cm} L{2.6cm}}
\br
Parameter           & Tolerance range$^{a}$\\
\mr
Despace             & $\pm$0.5 mm \\
Decenter	        & $\pm$0.15 mm \\
Tilt                & $\pm$1.2$^\prime$ (= $\pm$0.02$\degree$)\\
Focus (compensator)	& $\pm$0.5 mm \\
\br
\end{tabular} \\
$^{a}$\footnotesize{Tolerance ranges are common for all three mirrors (i.e., M1, M2, and M3).}
\end{table}

\begin{figure}[!ht]
\centering\includegraphics[width=7cm]{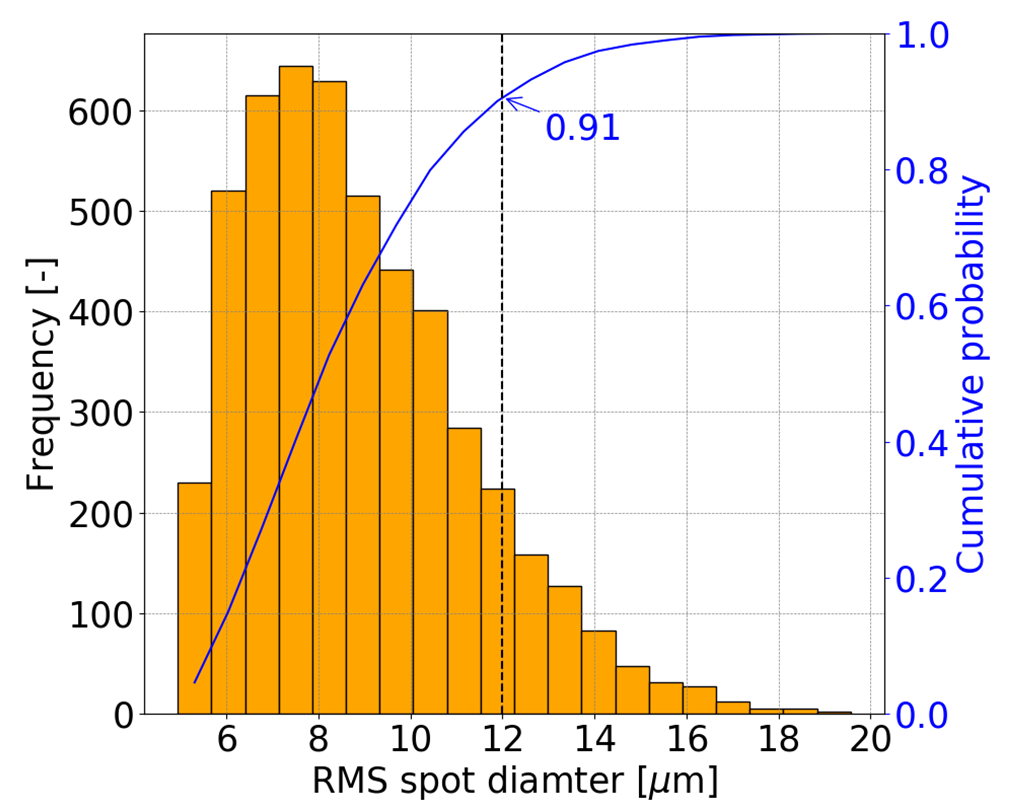}
\caption{\label{montecarlo}The Monte-Carlo simulation result confirming the optical performance of the LAF-TMS with realistic tolerances. The cumulative curve is shown in a blue solid line, and Nyquist sampling is indicated in a black dashed line.}
\end{figure}

Monte-Carlo simulation was evaluated with 5,000 trials. Criterion and reference wavelength are the same as those of the sensitivity analysis in Section~\ref{sec:sensitivity}. The statistical performance distribution of Monte-Carlo simulation is presented in Figure~\ref{montecarlo}. The blue solid line represents a cumulative curve, and the black dashed line indicates an optical requirement that corresponds to the Nyquist sampling with the visible CCD sensor format (12 $\mu$m, see Section~\ref{sec:opticdesign}). 

The Monte-Carlo analysis result indicates that the Nyquist sampling criteria at the telescope focal plane array corresponds to the 91 $\%$ cumulative probability. By considering the common precision manufacturing capabilities, tolerance limits are loose except for the tilt angles (Table~\ref{tab:tor}). In terms of risk management, when large errors occurred during fabrication and alignment processes, we implemented various realignment and compensation mechanisms to the optomechanical design described in Section~\ref{sec:optomechanics}.  

\section{Freeform aluminum mirror design and fabrication}
\label{sec:mirror}
Freeform mirror surfaces of the LAF-TMS can be expressed in the $\textit{xy}$ polynomial equations (\ref{eq2}) and (\ref{eq3}):
\small
\begin{eqnarray}
z = \frac{cr^2}{1+\sqrt{1-(1+k)c^2r^2}}+\sum_{j=2}^{66} C_jx^my^n, \label{eq2} \\
j = \frac{(m+n)^2+m+3n}{2}+1 \label{eq3}
\end{eqnarray}

\normalsize
In the above equations, $\textit{z}$ is the sag of the mirror surface parallel to the z-axis, $\textit{c}$ is the vertex curvature, $\textit{k}$ is the conic constant, $\textit{$C_{j}$}$ is the coefficient of the monomial $\textit{$x^{m}y^{n}$}$, and $\textit{$r^{2}$}$ is $\textit{$x^{2} + y^{2}$}$. Coefficients of odd power of $\textit{x}$ terms are zero since mirror surfaces are symmetric to $\textit{x}$ variables \citep{chang2019}. Designed mirror shape parameters are listed in Table~\ref{tab:mirrorshape}. The maximum sag deviations from conic surfaces ($\Delta z_{max}$) are 0.138, 0.245, and 0.179 mm for M1, M2, and M3, respectively.

\begin{table}[!ht]
\centering
\caption{Designed freeform mirror shape parameters of LAF-TMS}
\label{tab:mirrorshape}
\footnotesize
\begin{tabular}{L{1.3cm} L{1.7cm} L{1.7cm} L{1.7cm}}
\br
Para.$^{a}$	&	M1	&	M2	&	M3	\\
\mr
$\Delta z_{max}$    &   0.138 mm    &   0.245 mm    &    0.179 mm   \\
$\textit{c}$    	&	0 $mm^{-1}$	&	0 $mm^{-1}$ &	0 $mm^{-1}$	\\
$\textit{k}$	    &	-1	        &	-0.176	    &	-0.130	    \\
$\textit{$C_{4}$}$	&	-4.161E-04	&	-1.379E-03	&	-9.431E-04	\\
$\textit{$C_{6}$}$	&	-3.845E-04	&	-1.185E-03	&	-9.064E-04	\\
$\textit{$C_{8}$}$	&	9.642E-08	&	3.291E-07	&	1.132E-07	\\
$\textit{$C_{10}$}$	&	5.573E-10	&	-8.035E-07	&	-3.089E-08	\\
$\textit{$C_{11}$}$	&	1.010E-10	&	-2.672E-09	&	-9.864E-10	\\
$\textit{$C_{13}$}$	&	1.776E-10	&	-4.978E-09	&	-2.010E-09	\\
$\textit{$C_{15}$}$	&	1.006E-10	&	-2.013E-09	&	-8.828E-10	\\
$\textit{$C_{17}$}$	&	-1.549E-13	&	-2.356E-13	&	5.119E-13	\\
$\textit{$C_{19}$}$	&	-3.212E-13	&	-6.602E-12	&	2.919E-13	\\
$\textit{$C_{21}$}$	&	-8.606E-14	&	-6.873E-12	&	-4.598E-14	\\
$\textit{$C_{22}$}$	&	-1.819E-15	&	1.476E-13	&	-7.560E-16	\\
$\textit{$C_{24}$}$	&	-1.040E-15	&	-1.183E-13	&	-4.679E-15	\\
$\textit{$C_{26}$}$	&	-1.296E-15	&	-3.134E-13	&	-5.210E-15	\\
$\textit{$C_{28}$}$	&	1.558E-15	&	-2.488E-13	&	-1.130E-15	\\
$\textit{$C_{30}$}$	&	2.014E-17	&	-2.526E-16	&	-4.581E-17	\\
$\textit{$C_{32}$}$	&	4.583E-17	&	-4.882E-15	&	-2.779E-19	\\
$\textit{$C_{34}$}$	&	2.979E-17	&	-5.524E-15	&	-1.500E-18	\\
$\textit{$C_{36}$}$	&	-1.386E-18	&	-5.329E-16	&	-1.264E-17	\\
$\textit{$C_{37}$}$	&	2.496E-19	&	-1.234E-16	&	-2.237E-19	\\
$\textit{$C_{39}$}$	&	6.754E-20	&	2.437E-17	&	-6.129E-19	\\
$\textit{$C_{41}$}$	&	4.565E-19	&	1.918E-16	&	-9.466E-20	\\
$\textit{$C_{43}$}$	&	2.408E-19	&	2.789E-16	&	-2.451E-19	\\
$\textit{$C_{45}$}$	&	-2.213E-19	&	1.507E-16	&	-6.148E-20	\\
$\textit{$C_{47}$}$	&	-1.330E-21	&	5.540E-20	&	3.267E-21	\\
$\textit{$C_{49}$}$	&	-3.469E-21	&	2.064E-18	&	3.789E-21	\\
$\textit{$C_{51}$}$	&	-5.065E-21	&	3.426E-18	&	-2.837E-21	\\
$\textit{$C_{53}$}$	&	-1.606E-21	&	2.094E-18	&	1.494E-21	\\
$\textit{$C_{55}$}$	&	3.926E-23	&	8.833E-20	&	5.994E-22	\\
$\textit{$C_{56}$}$	&	-1.246E-23	&	2.932E-20	&	9.643E-24	\\
$\textit{$C_{58}$}$	&	7.462E-24	&	-6.612E-22	&	5.941E-23	\\
$\textit{$C_{60}$}$	&	-3.520E-23	&	-4.587E-20	&	-1.268E-23	\\
$\textit{$C_{62}$}$	&	-3.147E-23	&	-1.015E-19	&	2.687E-23	\\
$\textit{$C_{64}$}$	&	-1.502E-23	&	-9.517E-20	&	7.937E-25	\\
$\textit{$C_{66}$}$ &	1.174E-23	&	-3.389E-20	&	1.261E-24	\\
\br
\end{tabular} \\
$^{a}$\footnotesize{Parameters. Coefficients that are not listed in this table are zero.}
\end{table}

\begin{figure}[!ht]
\centering\includegraphics[width=8.3cm]{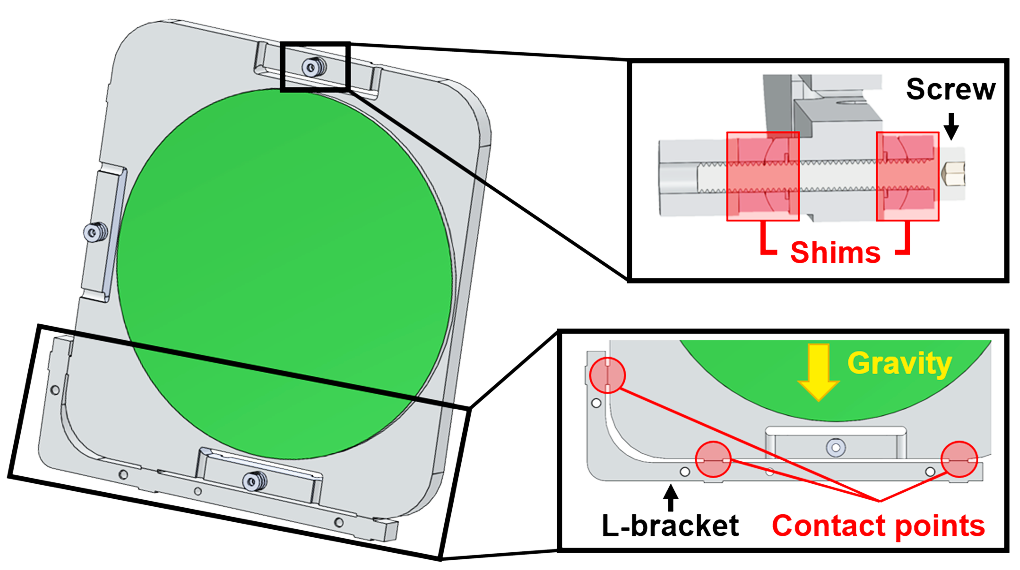}
\caption{\label{mechanicmirror}Mechanical design of the freeform aluminum mirror. (sub-figures) Precision made with the same aluminum material and alignment mechanisms are shown.}
\end{figure}

Off-axis mirrors are made of aluminum alloy 6061-T6 that conveniently mount on the same aluminum-based optomechanics. Applying the same material to optics and optomechanics increases thermal stability of the system. Figure~\ref{mechanicmirror} represents the mechanical design of the mirror and the alignment mechanism that allow the adjustment capability to compensate for any residual manufacturing errors beyond the tolerance limits. The thermal expansion slots and bent features in the mirror structure are designed to suppress thermal and mechanical stress on the reflecting surface. Mechanical deformations on the mirror surfaces due to the assembly process are minimized by optimizing these features. 

The 3-2-1 position principle is adapted to position the mirror \citep{trappey1990}. Shims are placed between mirrors and the mirror holder to adjust tilt and despace and to reduce stress from assembly process. The L-bracket is mounted underneath the mirror to support it. The mirror and the L-bracket meet at three contact points. We adjust x- and y-decenters by changing the thickness of the L-bracket. 

\begin{figure}[!ht]
\centering\includegraphics[width=8.3cm]{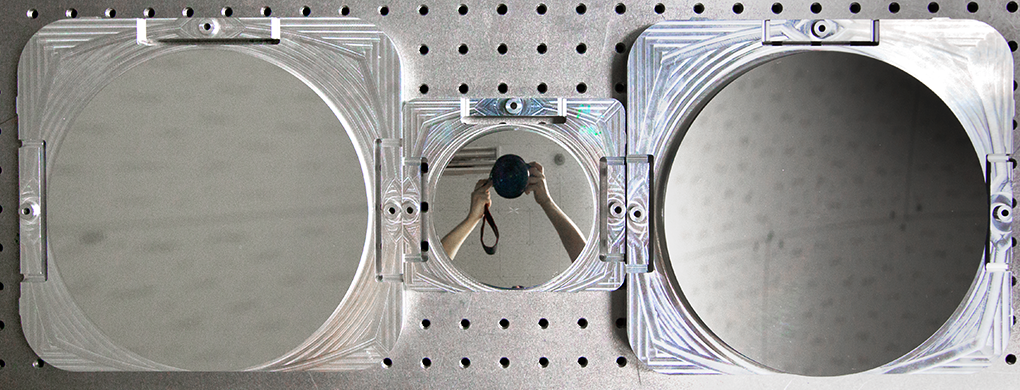}
\caption{\label{mirrors}Fabricated freeform 6061-T6 aluminum mirrors: (left) M1, (middle) M2, and (right) M3.}
\end{figure}

Precision manufacturing of the aluminum mirrors was produced through a Single Point Diamond Turning (SPDT) - Nickel plating - Polishing process. Nanotech 450 UPL and QED Q-FLEX 300 machines were used to fabricate freeform aluminum mirrors. The clear aperture size of mirrors is 180 mm for M1 and M3, and 86 mm for M2. Total dimensions of the mirror structure are 241 (L) $\times$ 222 (W) $\times$ 15 (H) mm for M1 and M3, and 125 (L) $\times$ 111 (W) $\times$ 14.5 (H) mm for M2 (Figure~\ref{mirrors}).

\begin{figure}[!ht]
\centering\includegraphics[width=8.3cm]{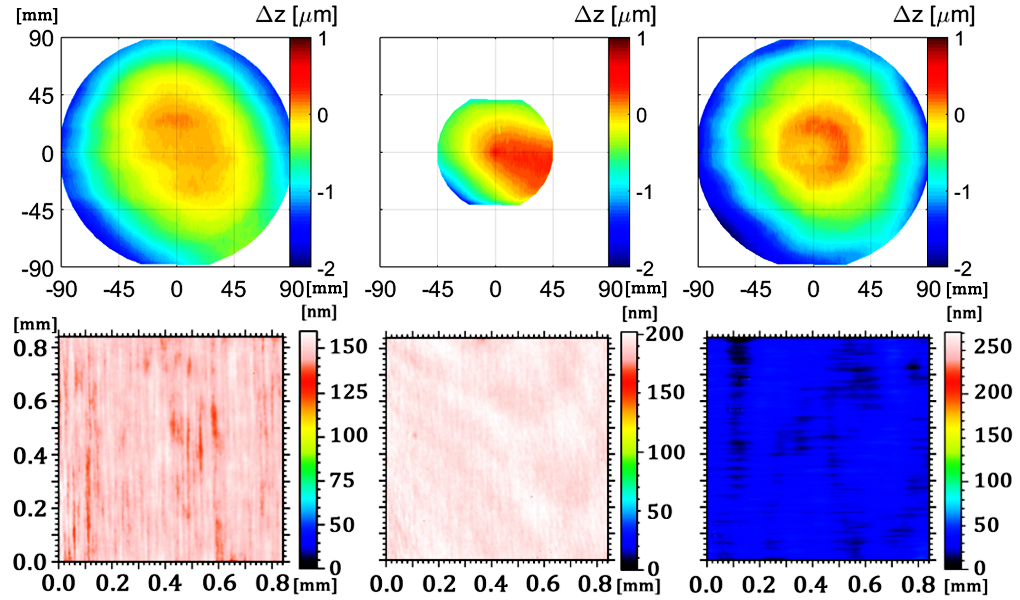}
\caption{\label{mirrorsurf}(top) Surface figure errors and (bottom) surface roughness maps: (left) M1, (middle) M2, and (right) M3 of the manufactured LAF-TMS prototype.}
\end{figure}

Figure~\ref{mirrorsurf} presents the measured surface shape error map (top) and micro roughness data (bottom) of the fabricated LAF-TMS mirrors. RMS surface figure errors are 0.403, 0.251, and 0.481 $\mu$m for M1, M2, and M3, respectively, when measured with the Ultrahigh Accurate 3-D Profilometer (UA3P, Panasonic) (Figure~\ref{mirrorsurf}, top).

Nickel plating on the aluminum mirror and polishing process significantly improved the surface finish, as shown in the micro roughness measurement data \citep{kim2015}. Magnetorheological Finishing (MRF) method reduces surface roughness (Ra) down to 2.3 nm for the M2 surface (see, Figure~\ref{mirrorsurf} middle). M3 shows higher surface roughness compared to those of M1 and M2, but it is still sufficiently good for science research in infrared wavelength. Surface measurement results are summarized in Table~\ref{tab:surferr}.

\begin{table}[!ht]
\centering
\caption{Surface shape errors and micro roughness of the as-manufactured LAF-TMS mirrors}
\label{tab:surferr}
\footnotesize
\begin{tabular}{L{3.2cm} L{1.1cm} L{1.1cm} L{1.1cm}}
\br
                & M1    & M2    & M3\\
\mr
RMS ($\mu$m)    & 0.40 & 0.25 & 0.48 \\
Peak-to-Valley ($\mu$m)	    & 1.6 & 1.4 & 2.2 \\
Ra (nm)         & 2.7 & 2.3 & 4.9\\
\br
\end{tabular} 
\end{table}

\section{Optomechanical design and simulation}
\label{sec:optomechanics}
\begin{figure}[!ht]
\centering\includegraphics[width=8.3cm]{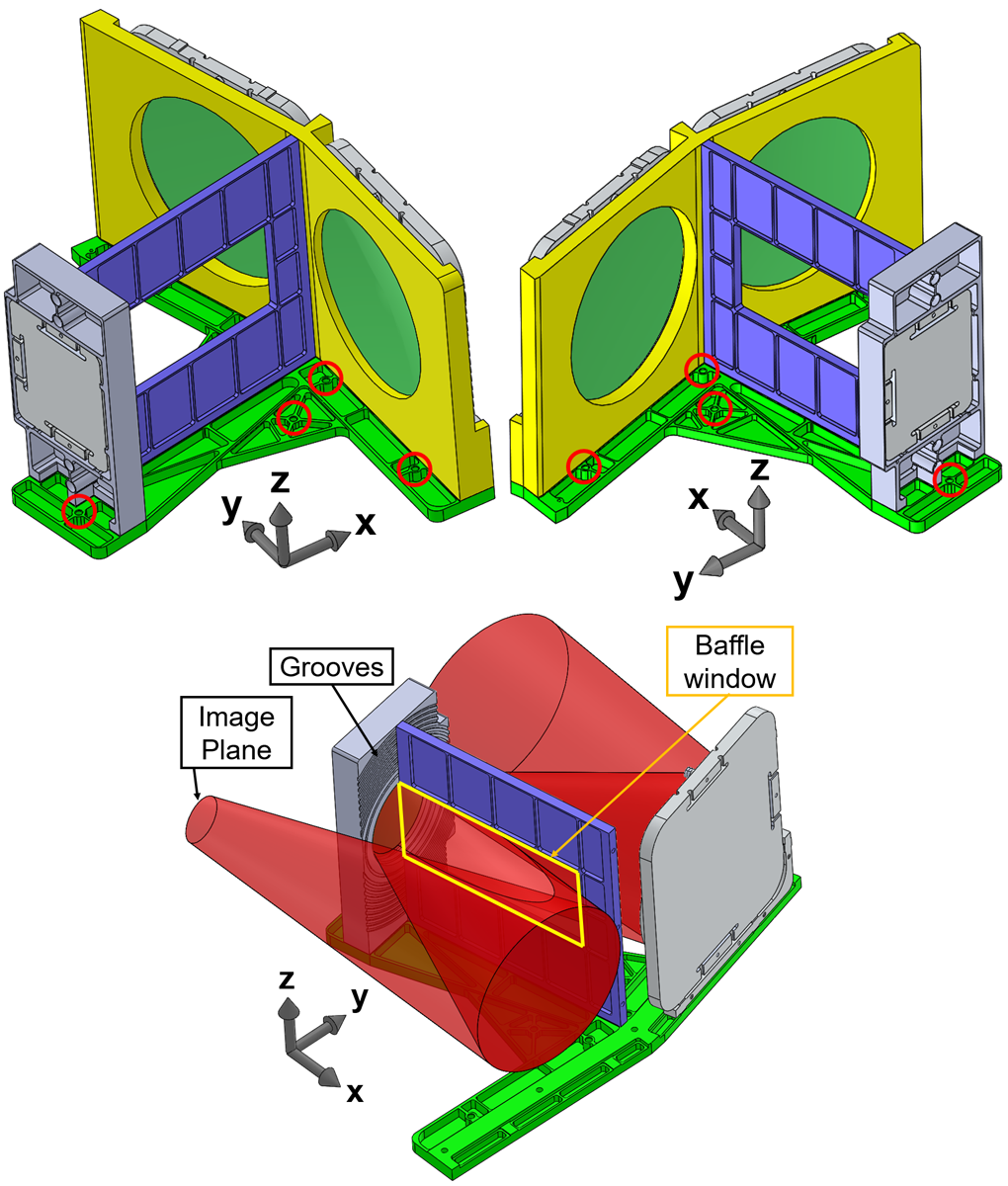}
\caption{\label{optomechanic}Optomechanical design of LAF-TMS. (top) Red circles are mounting positions of the base plate. (bottom) The optical path is illustrated in red. Groove features and the baffle window are indicated in the figure.}
\end{figure}

The optomechanical structure was designed to stably support mirrors at correct positions with the flexible modular structure approach. All parts are precisely assembled with pins and screws, and total dimensions are 351 (L) $\times$ 502 (W) $\times$ 266 (H) mm. All surfaces of the structure are anti-reflection black anodized. There are groove features on the M2 mirror holder surface and a baffle window to suppress stray light since a significant amount of light is reflected on the surfaces of the optomechanical structures (Figure~\ref{optomechanic}).

Satellites are exposed to various vibrational environments during critical launch events, such as left-off, wind and gust, stage separations, and etc. The stability of LAF-TMS is confirmed in vibration environments. Vibration environments were simulated using three analyses: quasi-static, harmonic, and random analysis. Modal analysis was also performed to calculate the system's natural frequency and mode shape \citep{abdelal2013}.

\subsection{Quasi-static analysis}
\label{sec:quasi-static}
Mass Acceleration Curve (MAC) has been adopted over many years for quasi-static analysis \citep{trubert1989}. Because it gives bound accelerations for each effective mass of the payload, quasi-static analysis with MAC is considered as the worst-case analysis. This analysis can be adapted to payload mass of less than 500 kg.

We used Space Shuttle and Inertial Upper Stage (STS/IUS) MAC to calculate quasi-static accelerations \citep{Changky2001}. The total mass of LAF-TMS is 9.47 kg, corresponding to 37.42-G, so we took the acceleration value of 40-G for all three axes. We fixed seven mounting positions and put accelerations on the same locations (red circles in Figure~\ref{optomechanic}). Quadratic tetrahedral 3D solid mesh elements were applied. The total number of nodes and elements are 144,404 and 79,128, respectively. All contact points and connections of parts are considered to be bonded.

Quasi-static analysis results are expressed in maximum von Mises stress, which derives Margin of Safety (MoS) with equation (\ref{eq4}) \citep{jeong2018}:
\begin{eqnarray}
MoS (\%) = \left[\left(\frac{\sigma_{yield}}{\sigma_{max} \times SF}\right)-1\right] \times 100 \% \label{eq4}
\end{eqnarray}

In equation (\ref{eq4}), $\sigma_{yield}$ is the yield stress of the material, and $\sigma_{max}$ is the maximum von Mises stress, which is the result of the simulation. Based on European Cooperation for Space Standardisation (ECSS) standards, safety factor (\textit{SF}) is 1.1 when using yield stress, and 1.25 for ultimate stress \citep{ecss2008a,ecss2008b,ecss2009,ecss2014}. We used aluminum alloy 6061-T6 for the entire system that has the yield stress of 275 MPa \citep{kaufman2000}.
\begin{table}[!ht]
\centering
\caption{Quasi-static loads and stress simulation results}
\label{tab:static}
\footnotesize
\begin{tabular}{L{1.5cm} L{2cm} L{1cm} L{1cm}}
\br
\multirow{2}{*}{Load-axis}       & Quasi-static load (G)    & \textit{$\sigma$\textsubscript{max}} (MPa)    & MoS ($\%$)\\
\mr
x    & 40 & 78.30  & 220 \\
y	 & 40 & 152.70 & 64 \\
z    & 40 & 34.86  & 620 \\
\br
\end{tabular} 
\end{table}

Table~\ref{tab:static} shows input quasi-static load, maximum von Mises stress, and MoS for each load axis. For all axes, LAF-TMS has positive MoS, indicating high stability of the telescope that overcomes the worst-case quasi-static environments.

\subsection{Modal analysis}
\begin{figure}[!ht]
\centering\includegraphics[width=8.3cm]{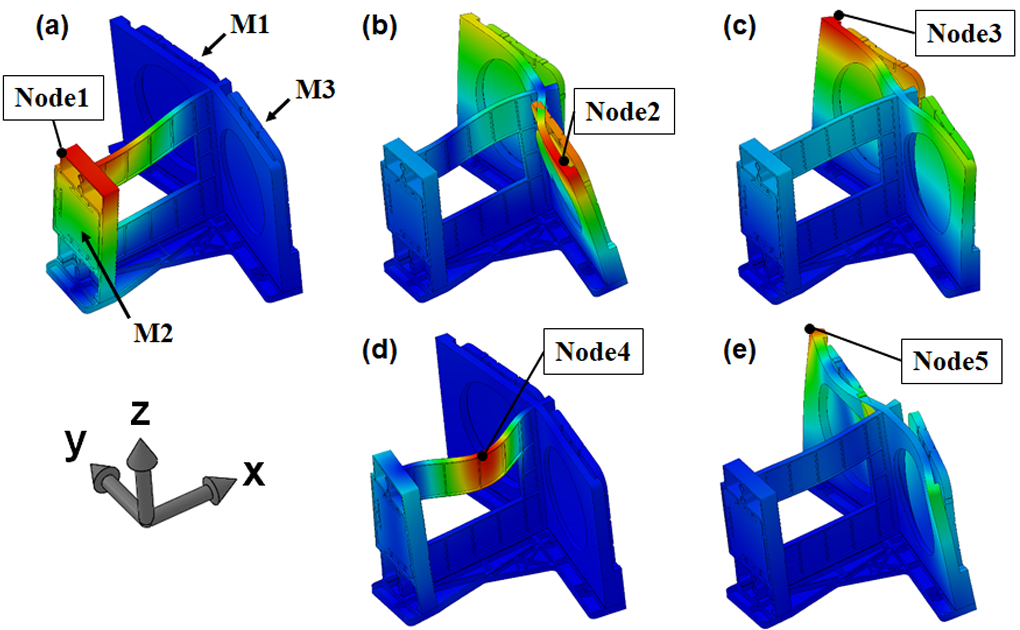}
\caption{\label{modeshape}Vibrational mode shapes, and harmonic and random vibration response nodes. Mechanical deformations in frequency (a) mode 1, (b) mode 2, (c) mode 3, (d) mode 4, and (e) mode 5 are illustrated. }
\end{figure}

Natural frequency and mode shape of LAF-TMS are examined with modal analysis. This analysis is the study of dynamic properties of system in frequency domain and helps optomechanics avoid exposure to vibration resonance \citep{ramesha2015}. Natural frequency is determined when structure shape, material, boundary conditions, and etc. are decided. The analysis results are summarized in Table~\ref{tab:modal} and Figure~\ref{modeshape}. 

\begin{table}[!ht]
\centering
\caption{Natural frequency of LAF-TMS}
\label{tab:modal}
\footnotesize
\begin{tabular}{P{2.8cm} P{3.8cm}}
\br
Frequency mode       & Natural frequency (Hz)  \\
\mr
1    & 121.57  \\
2	 & 157.88  \\
3    & 202.09  \\
4	 & 418.97  \\
5    & 455.50  \\
\br
\end{tabular} 
\end{table}

The fundamental frequency (frequency mode 1) is 121.57 Hz. Mode shape of frequency mode 1 shows that M2 might tilt in harsh vibration environments (Figure~\ref{modeshape}). We selected nodes for each vibration mode that specialize in measuring responses from harmonic and random vibrations.

\subsection{Harmonic and random analysis}
Harmonic and random analysis determine responses to sinusoidal and random loads, so it verifies whether LAF-TMS can survive these environments or not. We input vibration loads of the Souyz-2/Freget Launch system (see, Table~\ref{tab:vibinput}). The damping ratio of 0.02 modal damping is set for both analyses. Simulations are performed in x-, y-, and z-acceleration axes. All the other boundary conditions are the same as those of the quasi-static analysis in Section~\ref{sec:quasi-static}.

\begin{table*}[t]
\centering
\caption{Harmonic and random vibration qualification levels}
\label{tab:vibinput}
\footnotesize
\begin{tabular}{P{1.5cm} P{2.7cm} P{0.8cm} P{0.8cm} P{0.9cm} P{0.8cm} P{0.9cm} P{1.1cm} P{2cm}}
\br
\multirow{5}{*}{\shortstack{Harmonic \\ vibration}}       & \makecell{Frequency \\sub-range (Hz)} & \multicolumn{2}{c}{1 - 2} & \multicolumn{2}{c}{2 - 5} & \multicolumn{2}{c}{5 - 10} & 10 - 2000  \\
\cmidrule{2-9}
 &  \makecell{Vibration\\ accelerations (G) \\(G = 9.81 $m/s^{2}$)} & \multicolumn{2}{c}{\multirow{1}{*}{0.3 - 0.5}} & \multicolumn{2}{c}{\multirow{1}{*}{0.5}} & \multicolumn{2}{c}{\multirow{1}{*}{0.5 - 1.0}} & \multirow{1}{*}{1.0}  \\
\mr
\multirow{5}{*}{\shortstack{Random \\ vibration}} & \makecell{Frequency \\sub-range (Hz)} &  \makecell{20 - \\50} &  \makecell{50 -\\ 100} &  \makecell{100 - \\200} &  \makecell{200 - \\500} &  \makecell{500 - \\1000} &  \makecell{1000 - \\2000} & \multicolumn{1}{|c}{\makecell{Acceleration\\ RMS (G)}}  \\
\cmidrule{2-9}
& \makecell{Acceleration \\spectral density \\(ASD) ($G^{2}$/Hz)} & 0.02 & 0.02 & \makecell{0.02 -\\0.05} & 0.05 & \makecell{0.05 - \\0.025} & \makecell{0.025 - \\0.013} & \multicolumn{1}{|c}{7.42} \\
\br
\end{tabular} 
\end{table*}

\begin{figure}[!ht]
\centering\includegraphics[width=8.3cm]{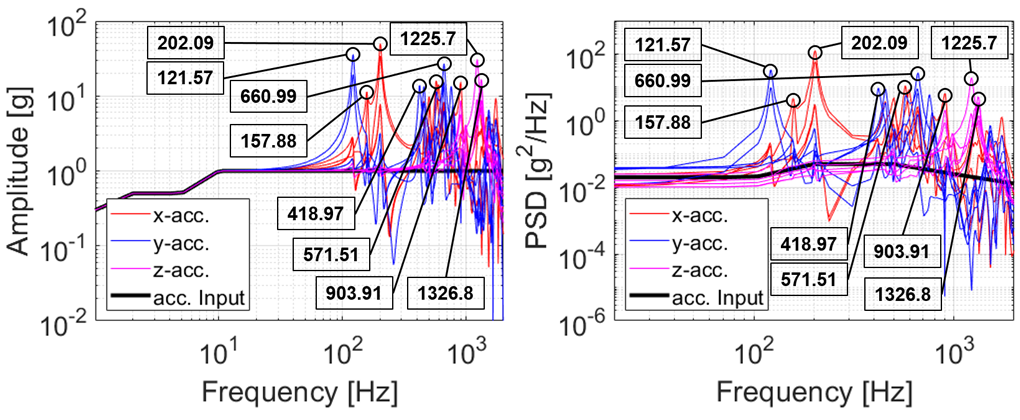}
\caption{\label{response}Response curves from harmonic (left) and random (right) analysis.}
\end{figure}

Figure~\ref{response} displays response curves from harmonic and random analysis. Each acceleration axis has five response curves that correspond to each of the response nodes (Figure~\ref{modeshape}). 

We identified eight dominant resonance frequencies, which are 121.57 Hz (mode 1), 157.88 Hz (mode 2), 202.09 Hz (mode 3), 418.97 Hz (mode 4), 571.51 Hz (mode 7), 660.99 Hz (mode 9), 903.91 Hz (mode 14), 1225.70 Hz (mode 18) and 1326.80 Hz (mode 20). Von Mises stress is calculated in these frequencies and the corresponding acceleration axes. 

\begin{table*}[t]
\centering
\caption{The maximum von Mises stress at nine dominant resonance frequencies from harmonic and random analysis}
\label{tab:vibresult}
\footnotesize
\begin{tabular}{P{4cm} P{2cm} P{2cm} P{2cm} P{2cm}}
\br
\multirow{3}{*}{\shortstack{Resonance frequency \\ (acceleration axis)}} & \multicolumn{2}{c}{Harmonic analysis}       & \multicolumn{2}{c}{Random analysis}  \\
\cmidrule{2-5}
 & \textit{$\sigma$\textsubscript{max}} (MPa) & MoS ($\%$) & \textit{$\sigma$\textsubscript{max}} (MPa) & MoS ($\%$) \\
\mr
121.57 (y)	&94.54	&165	&173.14	&45 \\
157.88 (x)	&14.31	&1653	&37.64	&567 \\
202.09 (x)	&56.09	&347	&166.94	&50 \\
418.97 (y)	&2.27	&10968	&9.67	&2496 \\
571.51 (x)	&2.77	&8961	&14.42	&1640 \\
660.99 (y)	&12.17	&1962	&50.26	&399 \\
903.91 (x)	&2.90	&8552	&16.11	&1458 \\
1225.70 (z)	&6.80	&3591	&27.41	&815 \\
1326.80 (z)	&7.10	&3433	&26.89	&833 \\
\br
\end{tabular} 
\end{table*}

Table~\ref{tab:vibresult} lists the dominant resonance frequencies. The maximum von Mises stress and MoS of each analysis are also presented. MoSs in all frequencies for both analyses are positive, indicating that LAF-TMS is safe from harmonic and random vibration environments of the launch system. 

\section{Optical performance verification}
\label{sec:perform}
\subsection{Optical alignment}
Optical alignment is performed before we verify and demonstrate the optical performance of LAF-TMS. The purpose of the alignment is to compensate for coordinate errors of fabricated optical components based on the coordinate measurement machine (CMM) measurements. By changing the thickness of shims and L-brackets, the positioning errors of mirrors can be relocated. Coordinates of optomechanical structure and mirrors were measured with the Dukin MHB CMM. 

Measured data points of each surface were fitted by using the least square fitting algorithm. All the fitted surfaces are compared with those of the designed nominal surfaces to calculate tilt and decenter of the mirrors. Table~\ref{tab:reopt} shows calculated tilt and decenter errors of each mirror before the alignment.

\begin{table}[!ht]
\centering
\caption{Tilt and decenter errors of the manufactured LAF-TMS mirrors measured by CMM}
\label{tab:reopt}
\footnotesize
\begin{tabular}{L{2.6cm} L{1.2cm} L{1.2cm} L{1.2cm}}
\br
							& M1 		& M2 	    	& M3  \\
\mr
$\alpha$-tilt ($\degree$)   & 0.071	& 0.192		&0.094 \\
$\beta$-tilt ($\degree$)	& 0.070	& -0.167	    &0.081 \\
x-decenter (mm)   			& 0.029	& 0.097		&0.040 \\
y-decenter (mm)	  			& 0.196	& -0.093   	&0.119 \\
z-decenter (mm)   			& -0.728	& 0.663		&-0.036 \\
\br
\end{tabular} 
\end{table}

Measured z-decenters for M1 and M2 are significantly large, potentially caused by compensation strategy during the mirror fabrication process \citep{zhang2015}. Tilt errors for all three mirrors are larger than the tolerance range ($\pm$0.02$\degree$) from the sum of mirror and optomechanical structure errors. We compensated for tilt and z-decenter by replacing shims of each mirror since tilt is the most critical parameter for optical performance. The final optical performance measurements were performed after re-positioning the mirrors.

\subsection{Point source test}
\begin{figure}[!ht]
\centering\includegraphics[width=8.3cm]{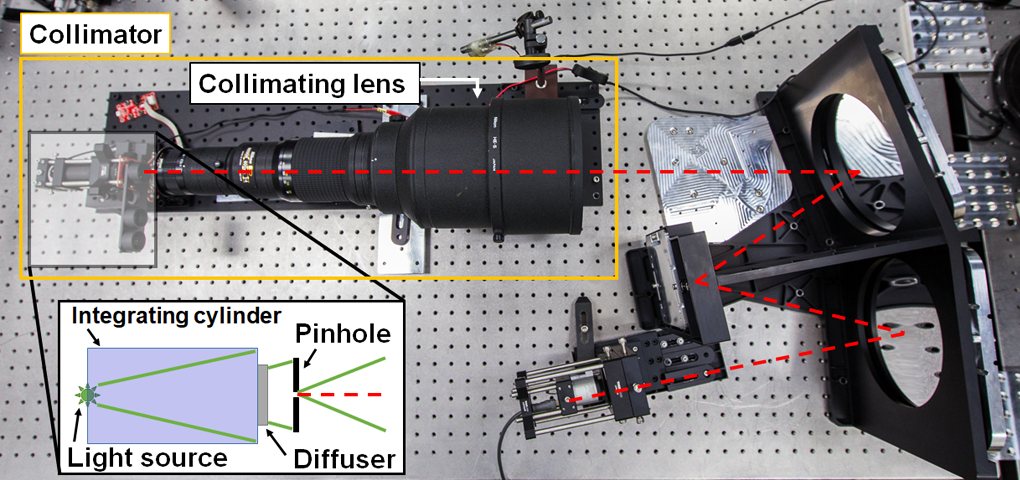}
\caption{\label{test}Optical test setup for point source imaging tests. The collimator is located on the left side, and LAF-TMS is installed on the right side. Optical axis ray is indicated in a red dashed line. (sub-figure) The optical layout of the collimator is illustrated in the black box.}
\end{figure}

For optical performance tests, we built the collimator system that can be tilted for full field tests. It consists of a white Light Emitting Diode (LED), integrating cylinder, diffuser, 5 $\mu$m-size pinhole, and high quality collimation lens. Figure~\ref{test} illustrates the layout of the imaging test setup, including the collimator and LAF-TMS prototype. We used a 3.75 $\mu$m pixel-sized CCD, QHY 5-II mono, to minimize measurement errors.

\begin{figure}[!ht]
\centering\includegraphics[width=8.3cm]{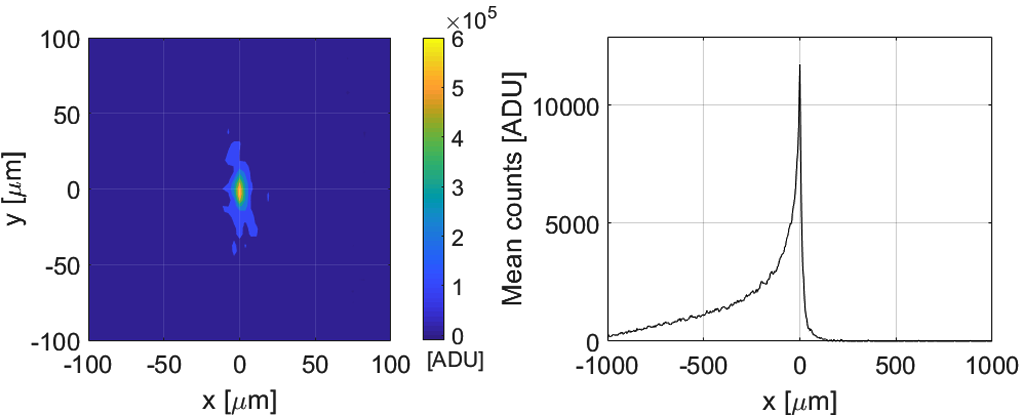}
\caption{\label{psf}(left) The contour plot of the point source image at the field center, and (right) its spot size.}
\end{figure}

The position of the sensor is controlled by linear stage with 0.01 mm accuracy. We subtracted dark frames and stacked 10 images to increase signal to noise ratios. Figure~\ref{psf} shows the contour plot of the point source image and its spot size. LAF-TMS has imaging performance with full width at half maximum (FWHM) of 37.3 $\mu$m (the right panel in Figure~\ref{psf}). Since optical performance of LAF-TMS targets the H2RG SCA infrared detector with the pixel size of 18 $\mu$m \citep{blank2012}, the spot size closely meets the Nyquist sampling theorem.

In addition to the center-field imaging performance verification, it is critical to perform the full FoV off-axis tests to confirm the significant strength of the linear-astigmatism-free optical design. To achieve large FoV, commercial DSLR camera (CMOS pixel size of 4.3 $\mu$m in the format of 22.3 $\times$ 14.9 mm or FoV of 2.55$\degree$ $\times$ 1.71$\degree$, Canon EOS 550D) with 2 $\times$ 2 binning is used as detector. In order to evaluate the image quality over the full FoV, point source images are obtained at 9 positions in different fields.

\begin{figure}[!ht]
\centering\includegraphics[width=8.3cm]{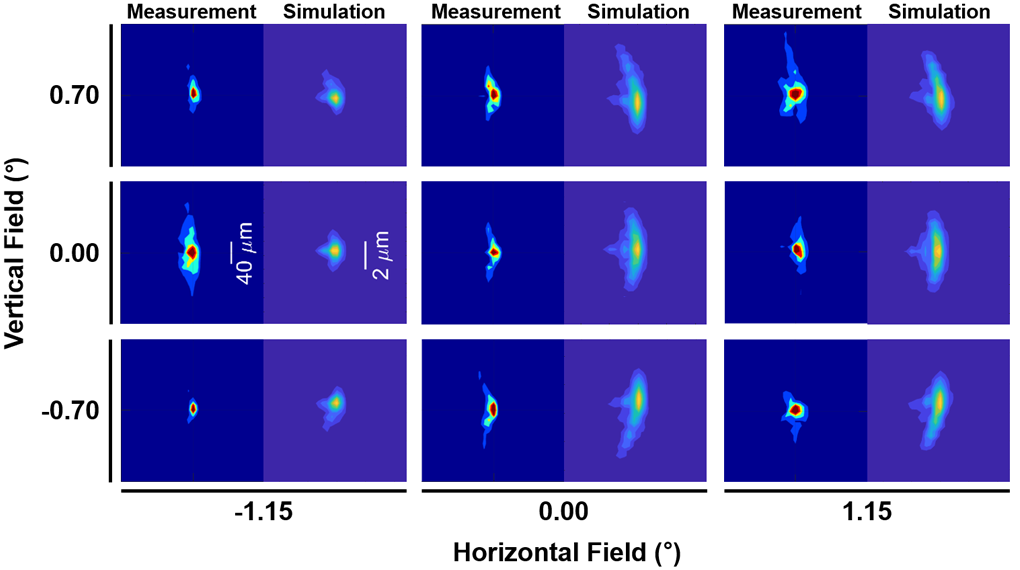}
\caption{\label{fullfield}(left) LAF-TMS prototype's full FoV test results in nine incident angles and (right) simulated ones at the same incident angles.}
\end{figure}

Figure~\ref{fullfield} shows full field imaging test and simulation results at the same field angle. The results successfully confirm that there are no dominant off-axis aberrations such as linear astigmatism. The comparison of measured data with the simulated point spread functions (PSF) yields some correlations between their shapes, while measured spots have 17.2 times larger size than the simulation, which sets the fundamental (i.e., ideal) performance limit. We suspect that the main reason for the large spot size is high surface figure errors of all three mirrors, which are 0.403, 0.251, and 0.481 $\mu$m for M1, M2, and M3, respectively. Research on the relationship between surface roughness and imaging performance revealed that scattered light from surface micro roughness of M3 (4.90 nm) may also amplify the spot size \citep{ingers1989,harvey2013}.

\section{Discussion and summary}
\label{sec:discuss}
We built a linear-astigmatism-free infrared telescope for satellite payloads. Optical design of the telescope is based on confocal off-axis three mirror reflective system. This design overcomes the limitations of available wavelength and FoV that are weaknesses of the traditional refractive and on-axis reflective telescopes, respectively. Also, there is no obscuration, scattering, and diffraction by optical components that appear in the on-axis system. Confocal off-axis design enables the telescope to have a simple, robust, and wide FoV telescope without any back-end corrective lenses. Therefore, LAF-TMS is a completely wavelength-independent optical system that provides enormous advantages for multi-band astronomical telescopes compared to classical off-axis design. Strength of the linear-astigmatism-free system as a multi-band telescope, which covers ultraviolet to infrared wavelength, is already verified \citep{hammar2019}.

Sensitivity and Monte-Carlo analysis show feasibility of building this system within general fabrication tolerances, but tilt errors must be carefully controlled as they are highly sensitive factors.

Freeform aluminum mirrors are designed with the 3-2-1 position principle including features that suppress thermal and mechanical stress from assembly torque. Surface RMS errors of the three mirrors are 0.403, 0.251, and 0.481 $\mu$m for M1, M2, and M3, respectively. The surface micro roughness is 2.70, 2.30, and 4.90 nm in the same order. 

Optomechanical structure is modularized and includes stray light suppression features. The fundamental frequency of the structure is 121.57 Hz, which is sufficiently high for a satellite payload. Quasi-static, harmonic, and random analysis were performed to identify and confirm survivability in vibration environments. In all vibration simulations, we confirmed positive margin of safeties in critical resonance frequencies. 

Mechanical fabrication and alignment errors were carefully measured with the CMM. By replacing spherical washers, we re-aligned $\alpha$- and $\beta$-tilt, and z-decenter errors. Point source measurements were performed at the image center as well as other FoV. The FWHM of the spot at the center is 37.3 $\mu$m, which closely meets to Nyquist sampling requirements for typical 18 $\mu$m pixel size infrared array detectors. Full field tests and simulations present similar image patterns, but the spot size is 17.2 times larger than measured ones. Based on sensitivity analysis, $\sim$0.4 $\mu$m surface RMS errors of each mirror can amplify spot size $\sim$10 times larger than the nominal spot size. Therefore, we conclude that high surface figure errors of M1, M2, and M3 are the main reason for the FWHM of measured PSF. The high surface roughness, especially for M3 (i.e. 4.90 nm), could cause the scattering effect that increases the spot size while retaining the image pattern. Other mirror substrate materials such as Zerodur or ULE (Ultra low expansion) can be considered in order to achieve better micro roughness values (e.g., \textless  2 nm RMS for a visible wavelength optical system application). It will improve the image contrast and overall throughput of the system by suppressing surface scattering at the cost of figuring-and-polishing expense and manufacturing time, which is usually defined by a computer-controlled optical surfacing process. However, optical test results show that figure and micro surface errors of LAF-TMS mirrors should be sufficiently acceptable for science research in infrared wavelength.

LAF-TMS is a prototype of next generation infrared telescopes for satellites, which covers wide FoV observation targets such as zodiacal light, integrated star light, transients, and other infrared astronomical sources. Linear-astigmatism-free off-axis reflective design is a versatile system that can be utilized not only for infrared observation but also for visible, sub-mm, and radio observations.

\bigskip
This work was supported by the National Research Foundation of Korea (NRF-2014M1A3A3A02034810 and 2017R1A3A3001362). Woojin Park and Sunwoo Lee were supported by by the BK21 Plus program through the NRF funded by the Ministry of Education of Korea. The development of the freeform mirrors were supported by Creative Convergence Research Project in the National Research Council of Science and Technology (NST) of Korea (CAP-15-01-KBSI). We thank the anonymous referees for their critical comments to improve this paper. We appreciate Mr. DongWook Kwak, Mr. WooRam Kim, and Mr. TaeHoon Kim in Green Optics for assisting the CMM measurements. We appreciate Dr. Young Ju Kim in Yunam Optics for coating and cleaning the freeform mirrors. We also thank Ms. Elaine S. Pak for proofreading this manuscript.

\newcommand{\newblock}{}
\bibliographystyle{aasjournal}
\bibliography{ref_tms}

\begin{thebibliography}{}
\expandafter\ifx\csname natexlab\endcsname\relax\def\natexlab#1{#1}\fi
\providecommand{\url}[1]{\href{#1}{#1}}
\providecommand{\dodoi}[1]{doi:~\href{http://doi.org/#1}{\nolinkurl{#1}}}
\providecommand{\doeprint}[1]{\href{http://ascl.net/#1}{\nolinkurl{http://ascl.net/#1}}}
\providecommand{\doarXiv}[1]{\href{https://arxiv.org/abs/#1}{\nolinkurl{https://arxiv.org/abs/#1}}}

\bibitem[{Abdelal {et~al.}(2013)Abdelal, Abuelfoutouh, \& Gad}]{abdelal2013}
Abdelal, G.~F., Abuelfoutouh, N., \& Gad, A.~H. 2013, in Finite Element
  Analysis for Satellite Structures: Applications to their design, manufacture
  and testing (Springer), 83--202

\bibitem[{Blank {et~al.}(2012)Blank, Anglin, Beletic, Bhargava, Bradley,
  Cabelli, Chen, Cooper, Demers, Eads, Farris, Lavelle, Luppino, Moore,
  Piquette, Ricardo, , Xu, \& Zandian}]{blank2012}
Blank, R., Anglin, S., Beletic, J.~W., {et~al.} 2012, Proc. SPIE, 8453, 845310

\bibitem[{Burge {et~al.}(2010)Burge, Benjamin, Dubin, Manuel, Novak, Oh,
  Valente, \& Zhao}]{burge2010}
Burge, J.~H., Benjamin, S., Dubin, M., {et~al.} 2010, Proc. SPIE, 7733, 77331J

\bibitem[{Chang(2001)}]{Changky2001}
Chang, K.~Y. 2001, ESASP, 468, 295

\bibitem[{Chang(2013)}]{chang2013}
Chang, S. 2013, Proc. SPIE, 8860, 88600U

\bibitem[{Chang(2015)}]{chang2015}
---. 2015, Journal of the Optical Society of America A, 32, 852

\bibitem[{Chang(2016)}]{chang2016}
---. 2016, Proc. SPIE, 9948, 99481B

\bibitem[{Chang(2019)}]{chang2019}
---. 2019, Optical Design and Fabrication (Freeform, OFT), FM2B, FM2B.6

\bibitem[{Chang {et~al.}(2006)Chang, Lee, Kim, Kim, Kim, Song, \&
  Park}]{chang2006}
Chang, S., Lee, J.~H., Kim, S.~P., {et~al.} 2006, ApOpt, 45, 484

\bibitem[{Chang \& Prata(2005)}]{chang2005}
Chang, S., \& Prata, A.~J. 2005, Journal of the Optical Society of America A,
  22, 2454

\bibitem[{Duval {et~al.}(2004)Duval, Irace, Mainzer, \& Wright}]{duval2004}
Duval, V.~G., Irace, W.~R., Mainzer, A.~K., \& Wright, E.~L. 2004, Proc. SPIE,
  5487, 101

\bibitem[{ESTEC(2008a)}]{ecss2008a}
ESTEC. 2008a, Space engineering Threaded fasteners handbook, ECSS-E-ST-32C
  Rev.1

\bibitem[{ESTEC(2008b)}]{ecss2008b}
---. 2008b, Space engineering Threaded fasteners handbook, ECSS-E-ST-32-03C

\bibitem[{ESTEC(2009)}]{ecss2009}
---. 2009, Space engineering Threaded fasteners handbook, ECSS-E-ST-32-10C
  Rev.1

\bibitem[{ESTEC(2014)}]{ecss2014}
---. 2014, Space engineering Threaded fasteners handbook, ECSS-E-ST-32-08C
  Rev.1

\bibitem[{Fazio {et~al.}(2004)Fazio, Hora, Allen, Ashby, Barmby, Deutsch,
  Huang, Kleiner, Marengo, Megeath, Melnick, Pahre, Patten, Polizotti, Smith,
  Taylor, Wang, Willner, Hoffmann, Pipher, Forrest, McMurty, McCreight,
  McKelvey, McMurray, Koch, Moseley, Arendt, Mentzell, Marx, Losch, Mayman,
  Eichhorn, Krebs, Jhabvala, Gezari, Fixsen, Flores, Shakoorzadeh, Jungo,
  Hakun, Workman, Karpati, Kichak, Whitley, Mann, Tollestrup, Eisenhardt,
  Stern, Gorjian, Bhattacharya, Carey, Nelson, Glaccum, Lacy, Lowrance, Laine,
  Reach, Stauffer, Surace, Wilson, Wright, Hoffman, Domingo, , \&
  Cohen}]{fazio2004}
Fazio, G.~G., Hora, J.~L., Allen, L.~E., {et~al.} 2004, ApJS, 154, 10

\bibitem[{{FLI}(2015)}]{fli}
{FLI}. 2015, {MicroLine Cameras}, https://www.flicamera.com/

\bibitem[{Funck \& Loosen(2010)}]{funck2010}
Funck, M.~C., \& Loosen, P. 2010, Proc. SPIE, 7652, 76521M

\bibitem[{Hammar {et~al.}(2019)Hammar, Park, Chang, Pak, Emrich, \&
  Stake}]{hammar2019}
Hammar, A., Park, W., Chang, S., {et~al.} 2019, ApOpt, 58, 1393

\bibitem[{Han {et~al.}(2014)Han, Lee, Jeong, Park, Moon, Park, Pyo, Kim, Park,
  LEE, SEON, NAM, CHA, PARK, PARK, YUK, REE, JIN, YANG, PARK, SHIN, SEO, RHEE,
  PARK, LEE, MURAKAMI, \& MATSUMOTO}]{han2014}
Han, W., Lee, D.-H., Jeong, W.-S., {et~al.} 2014, PASP, 126, 853

\bibitem[{Harvey(2013)}]{harvey2013}
Harvey, J.~E. 2013, OptEn, 52, 073110

\bibitem[{Houck {et~al.}(1984)Houck, Soifer, Neugebauer, Beichman, Aumann,
  Clegg, Gillett, Habing, Hauser, Low, Miley, Rowan-Robinson, \&
  Walker}]{houck1984}
Houck, J.~R., Soifer, B.~T., Neugebauer, G., {et~al.} 1984, ApJ, 278, L63

\bibitem[{{IAU}(2019)}]{minorplanet}
{IAU}. 2019, {Minor Planet Discoverers}, https://minorplanetcenter.net

\bibitem[{Ingers \& Breidne(1989)}]{ingers1989}
Ingers, J., \& Breidne, M. 1989, Proc. SPIE, 1029, 111

\bibitem[{Ishihara {et~al.}(2010)Ishihara, Onaka, Kataza, Salama, Alfageme,
  Cassatella, Cox, Garc\'{i}a-Lario, Stephenson, Cohen, Fujishiro, Fujiwara,
  Hasegawa, Ita, Kim, Matsuhara, Murakami, M\"{u}ller, Nakagawa, Ohyama, Oyabu,
  Pyo, Sakon, Shibai, Takita, Tanabe, Uemizu, Ueno, Usui, Wada, Watarai,
  Yamamura, , \& Yamauchi}]{ishihara2010}
Ishihara, D., Onaka, T., Kataza, H., {et~al.} 2010, A\&A, 514, 1

\bibitem[{Jeong {et~al.}(2018)Jeong, Kim, Chae, \& Jeon}]{jeong2018}
Jeong, G., Kim, J.~H., Chae, J., \& Jeon, H.-Y. 2018, Journal of Aerospace
  System Engineering, 12, 47

\bibitem[{Kaufman(2000)}]{kaufman2000}
Kaufman, J.~G. 2000, in Introduction to Aluminum Alloys and Tempers (ASM
  International), 39--76

\bibitem[{Kim {et~al.}(2010)Kim, Pak, Chang, Kim, Yang, Kim, Lee, \&
  Lee}]{kim2010}
Kim, S., Pak, S., Chang, S., {et~al.} 2010, JKAS, 43, 169

\bibitem[{Kim {et~al.}(2015)Kim, Chang, Pak, Lee, Jeong, Lee, Kim, Shin, \&
  Yoo}]{kim2015}
Kim, S., Chang, S., Pak, S., {et~al.} 2015, Applied Optics, 54, 10137

\bibitem[{Ku\'{s}(2017)}]{kus2017}
Ku\'{s}, A. 2017, ApOpt, 56, 9247

\bibitem[{Lee {et~al.}(2010)Lee, Hill, Marshall, Vattiat, \& DePoy}]{lee2010}
Lee, H., Hill, G.~J., Marshall, J.~L., Vattiat, B.~L., \& DePoy, D.~L. 2010,
  Proc. SPIE, 7735, 77353X

\bibitem[{Mainzer {et~al.}(2005)Mainzer, Eisenhardt, Wright, Liu, Irace,
  Heinrichsen, Cutri, \& Duval}]{mainzer2005}
Mainzer, A.~K., Eisenhardt, P., Wright, E.~L., {et~al.} 2005, Proc. SPIE, 5899,
  58990R

\bibitem[{Moon {et~al.}(2018)Moon, Park, Jeong, Lee, Ko, Lee, Park, Pyo, Park,
  Kim, Kim, Kim, Ko, Yu, Matsumoto, Chae, Shin, Takeyama, \&
  Enokuchi}]{moon2018}
Moon, B., Park, S.-J., Jeong, W.-S., {et~al.} 2018, Proc. SPIE, 10698, 106984R

\bibitem[{Neugebauer {et~al.}(1984)Neugebauer, Habing, van Duinen, Aumann,
  Baud, Beichman, Beintema, Boggess, Clegg, de~Jong, Emerson, Gautier, Gillett,
  Harris, Hauser, Houck, Jennings, Low, Marsden, Miley, Olnon, Pottasch,
  Raimond, Rowan-Robinson, Soifer, Walker, Wesselius, \&
  Young}]{neugebauer1984}
Neugebauer, G., Habing, H.~J., van Duinen, R., {et~al.} 1984, ApJ, 278, L1

\bibitem[{Onaka {et~al.}(2007)Onaka, Matsuhara, Wada, Fujishiro, Fujiwara,
  ISHIGAKI, ISHIHARA, ITA, KATAZA, KIM, MATSUMOTO, MURAKAMI, OHYAMA, OYABU,
  SAKON, TANAB\'{E}, TAKAGI, UEMIZU, UENO, USUI, WATARAI, COHEN, ENYA, OOTSUBO,
  PEARSON, TAKEYAMA, YAMAMURO, \& IKEDA}]{onaka2007}
Onaka, T., Matsuhara, H., Wada, T., {et~al.} 2007, Publ. Astron. Soc. Japan,
  59, S401

\bibitem[{Ramesha {et~al.}(2015)Ramesha, Abhijith, Singh, Raj, \&
  Naik}]{ramesha2015}
Ramesha, C.~M., Abhijith, K.~G., Singh, A., Raj, A., \& Naik, C.~S. 2015,
  IJETAE, 5, 323

\bibitem[{Ree {et~al.}(2010)Ree, Park, Moon, Cha, Park, Jeong, Lee, Nam, Park,
  Ka, Lee, Pyo, Lee, Rhee, Park, Lee, Matsumotod, Yang, \& Han}]{ree2010}
Ree, C.~H., Park, S.-J., Moon, B., {et~al.} 2010, Proc. SPIE, 7731, 77311X

\bibitem[{Trappey \& Liu(1990)}]{trappey1990}
Trappey, J.~C., \& Liu, C.~R. 1990, Int. J. Adv. Manuf. Technol., 5, 240

\bibitem[{Trubert(1989)}]{trubert1989}
Trubert, M. 1989, Mass acceleration curve for spacecraft structural design

\bibitem[{Wang {et~al.}(2013)Wang, Cheng, Wang, Hua, \& Jin}]{wang2013}
Wang, Q., Cheng, D., Wang, Y., Hua, H., \& Jin, G. 2013, ApOpt, 52, C88

\bibitem[{Werner(2012)}]{werner2012}
Werner, M. 2012, OptEn, 51, 011008

\bibitem[{Zhang {et~al.}(2015)Zhang, Zeng, Liu, \& Fang}]{zhang2015}
Zhang, X., Zeng, Z., Liu, X., \& Fang, F. 2015, OExpr, 23, 24800

\end{thebibliography}

\end{document}